# Pinned Magnetic

# Impurity Levels in Doped Quantum Dots

Nick S. Norberg,[1] Gustavo M. Dalpian,[2] James R. Chelikowsky,[2] and Daniel R. Gamelin[1,*]

[1]*Department of Chemistry, Box 351700, University of Washington, Seattle, WA 98195-1700*

[2]*Center for Computational Materials, Institute for Computational Engineering and Sciences, Departments of Physics and Chemical Engineering, University of Texas, Austin, Texas 78712*

*[*]Gamelin@chem.washington.edu*

**Abstract:** Spectroscopic data demonstrate that impurity D/A levels in doped semiconductor nanostructures are energetically pinned, resulting in variations in D/A binding energies with increasing quantum confinement. Using magnetic circular dichroism spectroscopy, the donor binding energies of Co2+ ions in colloidal ZnSe quantum dots have been measured as a function of quantum confinement and analyzed in conjunction with ab initio density functional theory calculations. The resulting experimental demonstration of pinned impurity levels in quantum dots has far-reaching implications for physical phenomena involving impurity-carrier interactions in doped semiconductor nanostructures, including in the emerging field of semiconductor spintronics where magnetic-dopant–carrier exchange interactions define the functionally relevant properties of diluted magnetic semiconductors.

Donor- and acceptor-type impurities play central roles in determining the physical properties of semiconductors. The energies of the impurity donor or acceptor (D/A) ionization levels relative to the respective band edges govern free-carrier concentrations, percolation thresholds, and metal-insulator transitions. The unique magnetic, magneto-electronic, and magneto-optical properties of diluted magnetic semiconductors (DMSs)[1] investigated for spintronics applications[2-4] also depend on the energies of the magnetic impurity D/A ionization levels relative to the respective band edges. As the dimensions of semiconductor devices are reduced to nanometer length scales, however, quantum confinement shifts the band edge energies relative to their bulk values, and changes in D/A binding energies may be anticipated. It is therefore of general importance to understand the impact of quantum confinement on D/A ionization energies in doped semiconductor nanostructures. For DMS nanostructures, it has previously been assumed that impurity D/A ionization energies are independent of quantum confinement of the host semiconductor.[5-7] Although such assumptions are substantiated theoretically in some cases,[8-10] they have not been verified directly by experimental determination of D/A ionization energies in quantum confined DMSs.

In this article we present spectroscopic and computational results for a series of $Co^{2+}$-doped ZnSe DMS quantum dots ($Co^{2+}$:ZnSe QDs) exhibiting various degrees of quantum confinement. Using magnetic circular dichroism (MCD) spectroscopy, a sub-bandgap charge-transfer (CT) transition was detected in $Co^{2+}$:ZnSe directly and unambiguously for the first time. This transition shifts to higher energy as the nanocrystal diameter is reduced, tracking quantitatively the destabilization of the conduction band induced by quantum confinement. Analysis yields the firm experimental conclusion that the $Co^{2+}$ ionization levels are pinned in energy as the ZnSe band energies shift with increasing quantum confinement. Density functional theory (DFT) calculations support this analysis and yield quantitative descriptions of the CT excited-state wavefunctions. The direct observation of pinned impurity levels has far-reaching implications for numerous physical phenomena involving impurity-carrier interactions in doped semiconductor nanostructures, including carrier-mediated ferromagnetism, magnetic polaron nucleation, and metal-insulator transitions in nanoscale DMSs relevant to spintronics.

Fig. 1A shows 300 K electronic absorption spectra of colloidal 5.6 nm diameter 0.77% $Co^{2+}$:ZnSe QDs (i.e. $Zn_{0.9923}Co_{0.0077}Se$). The first excitonic absorption feature at 2.98 eV is blue shifted from its energy in bulk ZnSe at 300 K (2.69 eV).[11] A further blue shift is observed upon lowering the temperature from 300 to 5 K (Fig. 1B), consistent with the expected temperature dependence of the band gap energies of bulk and nanocrystalline ZnSe.[12,13] When the $Co^{2+}$:ZnSe QD suspension is concentrated, another feature is observed at 1.67 eV that is readily identified as the $^4A_2(F) \rightarrow \ ^4T_1(P)$ ligand-field transition of $Co^{2+}$ substitutionally doped in ZnSe (Fig. 1A). This ligand-field transition's energy and band shape are sensitive to the $Co^{2+}$ coordination environment as described by ligand-field theory, and its similarity to the corresponding bulk spectrum (dashed line in Fig. 1A)[14] confirms the homogeneous $Co^{2+}$ speciation as a substitutional dopant within the ZnSe QD lattice. There is no evidence of any other $Co^{2+}$ species.

Fig. 1B shows the 5 K electronic absorption spectra of three $Co^{2+}$:ZnSe QD samples with average nanocrystal diameters of 5.6, 4.6, and 4.1 nm (0.77, 1.30, 0.61% $Co^{2+}$, respectively). Due to quantum confinement, the excitonic transition shifts to higher energy with decreasing nanocrystal diameter across this series. Fig. 1C shows 5 K, 6 T MCD spectra of the same three $Co^{2+}$:ZnSe QD samples. In each MCD spectrum, three features are observed: (*i*) the $^4A_2(F) \rightarrow \ ^4T_1(P)$ transition at 1.67 eV, (*ii*) a negative feature at ca. 2.85 eV, and (*iii*) an intense derivative-shaped feature centered at ca. 3.10 eV that is associated with the first excitonic transition of ZnSe. The MCD signal intensities for all three features in each sample follow the saturation behavior expected from the S = 3/2 Brillouin function calculated using g = 2.27 from bulk

Co$^{2+}$:ZnSe,[15] indicating that all MCD intensities originate from the same substitutionally doped Co$^{2+}$ chromophore. Analysis of the excitonic MCD intensity reveals giant Zeeman splittings in these Co$^{2+}$:ZnSe QDs, as described elsewhere.[16] Like the excitonic transition, the sub-bandgap transition shifts to higher energy with decreasing nanocrystal size, as shown in Fig. 2A on an expanded energy scale. In contrast, there is no detectable dependence of the ligand field transition energy on particle size (Fig. 1C).

The negative MCD band at 2.85 eV in the Co$^{2+}$:ZnSe QDs is absent from the MCD spectra of ZnSe or Mn$^{2+}$:ZnSe QDs,[17] confirming its association with Co$^{2+}$. This transition can be assigned as a CT transition involving Co$^{2+}$ and one of the semiconductor bands because of its broad shape, which is typical of CT transitions, and because no spin-allowed ligand-field transitions occur in this energy range. In general, two types of CT transitions may be anticipated in DMSs: (i) donor-type promotion of an electron from the dopant to the conduction band, referred to here as a metal-to-ligand CT (ML$_{CB}$CT) transition, and (ii) acceptor-type promotion of an electron from the valence band to the dopant, referred to here as a ligand-to-metal CT (L$_{VB}$MCT) transition. To differentiate between the two possible assignments, this transition's shift in energy with particle size was compared with the expected shifts of the conduction- and valence-band edges for the same particles.

The CT energies in each spectrum were determined by Gaussian fitting of the 5 K MCD and absorption spectra. Four Gaussians were used to account for the four allowed excitonic transitions,[1] and one Gaussian was used to account for the CT transition. The band-shape parameters for the excitonic MCD signals were obtained from the low temperature absorption spectra. Fig. 2B plots the CT onset energies versus peak excitonic energies for the three Co$^{2+}$:ZnSe QD samples (solid circles). The peak excitonic energies are also plotted for comparison (solid diamonds). The CT data in Fig. 2B fit very well to a straight line with a slope of $\Delta E_{CT}/\Delta E_{EXC} = 0.80 \pm 0.03$. Extrapolation of this best fit line from the nanocrystals to the bulk excitonic energy of 2.804 eV[18] predicts a CT onset energy of $2.51 \pm 0.07$ eV (Fig. 2B, open circle) for bulk $Co^{2+}$:ZnSe. This value is within experimental error of most low-temperature $ML_{CB}CT$ energies estimated previously for bulk $Co^{2+}$:ZnSe[19-22] (Fig. 2B, red bars). The CT transition observed by MCD is thus the $ML_{CB}CT$ transition identified in bulk $Co^{2+}$:ZnSe, but its energy has been shifted relative to bulk due to quantum confinement.

To analyze $\Delta E_{CT}/\Delta E_{EXC}$, the effective mass approximation (Eq. 1)[23] describing $\Delta E_{EXC}$ versus particle radius ($r$) was used, where $m_e*$ and $m_h*$ are the effective masses of the electron and hole for the semiconductor, $e$ is the electron charge, and $\varepsilon$ is the semiconductor bulk dielectric constant.

$$\Delta E_{EXC} = \frac{h^2}{8r^2}\left[\frac{1}{m_e^*} + \frac{1}{m_h^*}\right] - \frac{1.8e^2}{\varepsilon r} \tag{1}$$

From Eq. 1, $\Delta E_{CB}$ and $\Delta E_{VB}$ can be estimated relative to $\Delta E_{EXC}$ as in Eqs. 2a,b.[13]

$$\frac{\Delta E_{CB}}{\Delta E_{EXC}} \approx \frac{m_e^{*-1}}{m_e^{*-1} + m_h^{*-1}} \tag{2a}$$

$$\frac{\Delta E_{VB}}{\Delta E_{EXC}} \approx \frac{m_h^{*-1}}{m_e^{*-1} + m_h^{*-1}} \tag{2b}$$

From Eq. 2 and the effective masses of the electron and hole in ZnSe ($m_e* = 0.16$ and $m_h* = 0.74$),[18] $\Delta E_{CB}/\Delta E_{EXC} = 0.82$ and $\Delta E_{VB}/\Delta E_{EXC} = 0.18$. Comparison with the experimental ratio ($\Delta E_{CT}/\Delta E_{EXC} = 0.80 \pm 0.03$, Fig. 2B) supports assignment of this band as the $ML_{CB}CT$ transition.

The quantitative agreement between experimental $\Delta E_{CT}/\Delta E_{EXC}$ and predicted $\Delta E_{CB}/\Delta E_{EXC}$ ratios allows two important conclusions to be drawn. First, it indicates that the photogenerated electron arising from the $ML_{CB}CT$ transition is truly delocalized in the ZnSe conduction band (*i.e.*, $Co^{2+} \rightarrow Co^{3+} + e_{CB}^-$). An excited electron bound at the $Co^{3+}$ by strong Coulombic interactions would have a larger effective mass than the free conduction band electron, and consequently a smaller value of $\Delta E_{CT}/\Delta E_{EXC}$ would be observed in Fig. 2B. Second, the data demonstrate that the energies of the $Co^{2+}$ levels are pinned, independent of quantum confinement of the host ZnSe. These conclusions are summarized schematically in Fig. 2C.

The electronic structures of quantum confined $Co^{2+}$:ZnSe nanocrystals were further investigated using *ab initio* methods based on DFT using norm-conserving pseudopotentials within the local spin-density approximation (LSDA) in a real space approach.[10] Clusters of 35, 87, 147, and 297 atoms excluding surface-terminating H (diameters = 1.11, 1.45, 1.76, and 2.27 nm) were investigated, each with one $Co^{2+}$ at the origin. Although LSDA-DFT calculations typically underestimate band gap energies (e.g., 1.47 (DFT) vs 2.82 eV (experiment) for bulk ZnSe[18]), the calculations provide a valuable framework for examination of the sources of the trends in photoionization energies observed in quantum confined $Co^{2+}$:ZnSe nanocrystals.

To calculate CT transition energies and photoexcited hole and electron wavefunctions by DFT, a constrained LSDA approach was used in which the occupations of the conduction band and impurity levels were fixed. For comparison with the experimental data, Fig. 2D plots the calculated D/A transition energies on the same axes as the experimental results of Fig. 2C. Consistent with experiment, the calculated $ML_{CB}CT$ transition energies shift with particle size and the entire shift derives from quantum confinement effects on the conduction band. The DFT calculations thus support the experimental conclusion from Fig. 2B that the $Co^{2+}$ donor level is energetically pinned in quantum confined $Co^{2+}$:ZnSe. The calculated electron and hole wavefunctions in the $ML_{CB}CT$ excited state clearly illustrate the origin of this relationship. Fig. 3A shows the calculated probability density for the photoexcited electron of the 1.47 nm diameter nanocrystal. This electron is delocalized over the entire nanocrystal in an *s*-like orbital ($A_1$ irreducible representation in the $T_d$ point group of the calculated structure) having little (~0.5%) cobalt 4*s* character and no cobalt 3*d* character. The absence of cobalt *d*-orbital character is partially due to symmetry restrictions of the calculations. In contrast with the photoexcited electron, the photogenerated hole probability density (Fig. 3B) is highly localized at the impurity, having predominantly (~70%) cobalt $3d(e(\downarrow))$ character. This description does not change qualitatively over the entire range of particle diameters investigated. In the $ML_{CB}CT$ excited state, the highly localized hole (*d* orbital) is not influenced by particle size, whereas the delocalized $e^-_{CB}$ is. The $ML_{CB}CT$ transition energy thus depends on particle size in the same way as the CB does, and hence the experimental slope in Fig. 2B is the same as that anticipated from Eq. 2a. The electron and hole wavefunctions calculated by DFT thus support the description of this excitation as $Co^{2+} \rightarrow Co^{3+} + e^-_{CB}$ determined experimentally from the slope of $\Delta E_{CT}/\Delta E_{EXC}$ (Fig. 2B).

The DFT results also allow description of the $L_{VB}MCT$ excited state, which is not observed in Fig. 1 due to overlap with the intense ZnSe band gap absorption.[22] Fig. 2D plots the $L_{VB}MCT$ transition energies calculated for $Co^{2+}$:ZnSe QDs (dashed arrow). Like the $ML_{CB}CT$ energies, the calculated $L_{VB}MCT$ energies shift with quantum confinement. This shift is entirely attributable to the effect of quantum confinement on the VB, and the $Co^{2+}$ acceptor level is energetically pinned.

This experimental demonstration of pinned impurity levels in quantum confined semiconductor nanocrystals has important general implications for the properties of doped semiconductor nanostructures. In the field of spintronics, for example, carriers generated by donors or acceptors are necessary for ferromagnetism in many DMSs,[3,4] but the data here indicate that the depths of these impurity levels will change with increasing quantum confinement. To illustrate, the $Co^{3+}$ formed after $ML_{CB}CT$ excitation can itself be described as a deep acceptor (i.e., $Co^{2+}$ with a strongly bound VB hole ($h^+_{VB}$)).[24] Fig. 2C,D shows that the binding energy ($E_b$) of the hole changes as the VB shifts due to quantum confinement. This energy difference determines the effective Bohr radius of $h_{VB}^+$ as described by Eq. 3.[25]

$$r_B = \frac{\hbar}{\sqrt{2m_h^* E_b}} \tag{3}$$

From the VB shift predicted by Eq. 2b and the data in Fig. 2, $r_B$ of the $Co^{2+}$-bound $h_{VB}^+$ shrinks from 0.50 nm ($E_b$ = 250 meV) in bulk to 0.44 nm ($E_b$ = 336 meV) in 4.1 nm diameter $Co^{2+}$:ZnSe nanocrystals (Fig. 4A), a 12% decrease. This trend is also found in the DFT results of Fig. 3C, which compares density-of-states diagrams for the $ML_{CB}CT$ excited states of 1.11 and 2.27 nm diameter $Co^{2+}$:ZnSe nanocrystals. As the nanocrystal diameter is reduced, the cobalt $e(\downarrow)$ orbitals are separated energetically from the valence band edge and move deeper into the gap. Since the $ML_{CB}CT$ excited-state hole resides in $e(\downarrow)$, this increased separation corresponds directly to increased localization of the hole at the cobalt due to quantum confinement.

The impact of quantum confinement will be greater for shallower donors or acceptors and in semiconductors with smaller $m_e^*$ and $m_h^*$. For example, the isovalent dopant $Mn^{3+}$ in bulk $Ga_{1-x}Mn_xAs$ acts as a shallow acceptor, forming $Mn^{2+}$ with a bound hole having $r_B$ = 0.78 nm ($E_b$ = 113 meV).[26] This large Bohr radius is essential for carrier-mediated ferromagnetism in $Ga_{1-x}Mn_xAs$ and it determines the critical manganese acceptor concentration ($n_{crit}$) necessary for the metal-insulator transition, as described by Eq. 4.[27]

$$\left(n_{crit}\right)^{1/3} r_B = 0.26 \tag{4}$$

Using $m_h^*$ = 0.56, $\varepsilon$ = 10.66 for GaAs[26] and Eqs. 1,2b to determine the VB energy shift, $r_B$ is predicted to decrease substantially with quantum confinement (Fig. 4A), dropping to ca. 56% of its bulk value in 3 nm diameter QDs, and $n_{crit}$ increases concomitantly (Fig. 4B). At 4.8 nm diameter, $r_B \approx 0.62$ nm ($E_b \approx 180$ meV) and $n_{crit}$ has doubled relative to bulk $Ga_{1-x}Mn_xAs$. Although metallic conductivity is not required for ferromagnetism in $Ga_{1-x}Mn_xAs$, the predicted decrease of $r_B$ in $Ga_{1-x}Mn_xAs$ nanostructures will reduce the number of manganese ions that interact with a given hole, which in turn can reduce the ferromagnetic Curie temperature or even destroy ferromagnetism completely.[3,4] Alternatively, in situations where the D/A levels reside outside of the bulk band gap, quantum confinement can reduce the energy difference between these and the relevant band edges, thereby increasing dopant-carrier hybridization.[9] Energetic pinning of magnetic impurity levels in quantum confined DMSs may thus have significant implications for physical properties that rely on carriers, including ferromagnetic ordering and metal-insulator transitions.

More generally, knowledge of impurity-carrier binding energies will be critical for the development of future information processing technologies based on doped quantum wells, quantum wires, or quantum dots. The direct observation here of pinned impurity levels in quantum confined $Co^{2+}$:ZnSe implies that universal alignment rules[28-30] widely used to interpret the D/A ionization energies of doped bulk semiconductors are also applicable to doped semiconductor nanostructures. The binding energies of numerous electronic donors or acceptors may thus be estimated reliably in a wide range of nanoscale semiconductors from knowledge of their bulk binding energies and the effects of quantum confinement on the relevant band-edge potentials of the host semiconductor.

**Methods**

For preparation of $Co^{2+}$:ZnSe QDs, a cation solution of $Zn(OAc)_2 \cdot 2H_2O$ and $Co(OAc)_2 \cdot 4H_2O$ (0.4 mmol total) dissolved in oleic acid (1.2 mmol), HDA (1.6 mmol), and ODE (16 g) was degassed with bubbling $N_2$ at 120 °C for about one hour. Elemental selenium (2.4 mmol) was dissolved in TBP (4 mmol) and ODE (1.2 g) in a glove box and stored under $N_2$. The cation solution was heated to 310 °C and the Se solution was rapidly injected using a gas-tight syringe. The temperature of the reaction solution dropped and was subsequently stabilized at 305 °C, where it was kept until the desired nanocrystal size was reached. The $Co^{2+}$:ZnSe nanocrystals were precipitated with a mixture of toluene and ethanol, centrifuged, and resuspended in toluene several times to wash away precursor material. These washed nanocrystals could be redispersed toluene or other nonpolar organic solvents. A detailed description of the synthesis and characterization of these $Co^{2+}$:ZnSe nanocrystals will be published separately.[16]

300 K electronic absorption spectra of colloidal $Co^{2+}$:ZnSe QDs were collected using 1.0 cm cuvettes and a Cary 500 (Varian) spectrophotometer. Low temperature electronic absorption and magnetic circular dichroism (MCD) spectra were collected on $Co^{2+}$:ZnSe nanocrystals drop-coated onto quartz disks using an Aviv 40DS spectropolarimeter with a sample compartment modified to house a high-field superconducting magneto-optical cryostat (CryoIndustries SMC-1659M-OVT) positioned in the Faraday configuration. Absorption intensities were measured in single-beam mode. MCD intensities were measured as the absorbance difference ($\Delta A = A_L - A_R$, where L and R refer to left and right circularly polarized photons). CT onset energies were determined from Gaussian fitting of MCD and absorption spectra as described previously.[16] Briefly, four Gaussians were used to account for the four allowed excitonic transitions in the MCD, and one to account for the CT transition. The bandshape parameters for the excitonic MCD signals were determined independently from simultaneous absorption measurements, and their theoretical intensity ratios were assumed. For comparison with literature data, the CT onset energies (approximate electronic origins) are reported in Fig. 2B. Multiple specific fitting approaches were explored. Each yielded a straight line as shown in Fig. 2B, but with vertical offsets depending on the specific approach used. This systematic uncertainty is accounted for in the error bars on the extrapolated bulk CT energy in Fig. 2B. Nanocrystal sizes were estimated from room temperature excitonic absorption energies using a previously reported relationship,[31] and in specific cases these sizes were verified by TEM measurements and X-ray diffraction linewidth analysis.

The ZnSe valence- and conduction-band energy shifts due to quantum confinement (Fig. 2C) were calculated from Eqs. 2a,b and the relationship between ZnSe nanocrystal diameter and excitonic energy reported previously.[31] These energies in Fig. 2C are plotted relative to vacuum based on the experimental valence-band ionization energy for bulk ZnSe.[32] The GaAs valence-band shift vs particle diameter used to determine $r_B$ and $n_c$ in Figs. 4A,B was estimated from Eq. 1 by considering only $m_h*$ (i.e. neglecting $m_e*$) and only half of the Coulombic term.

The electronic structures of $Co^{2+}$:ZnSe nanocrystals were studied using density functional theory and the pseudopotentials method[33] within the local spin density approximation (LSDA). The Kohn-Sham equations were solved on a real space grid using a higher-order finite difference method.[34] A grid spacing of 0.35 a.u. and a separation of at least 6.0 a.u. between the outermost passivating atoms and the spherical boundary were used throughout. The Zn and Se pseudopotentials were generated in a similar way as in ref. 10, while for Co we used a cutoff radius of 2.0 a.u. for *s* and *p* orbitals and 2.2 a.u. for *d* orbitals.

A tetrahedrally coordinated $Co^{2+}$ ion was substituted at a $Zn^{2+}$ site in the center of each nanocrystal. We considered four $Co^{2+}$-doped ZnSe QDs: $Zn_{18}CoSe_{16}$, $Zn_{42}CoSe_{44}$, $Zn_{78}CoSe_{68}$, and $Zn_{140}CoSe_{152}$. These cases correspond to effective nanocrystal diameters of 1.11, 1.45, 1.76, and 2.27 nm, respectively, and effective $Co^{2+}$ cation mole fractions of 5.5, 2.3, 1.4, and 0.7%, respectively. All of the surface atoms are passivated with hypothetical H atoms to maintain charge neutrality.

Transition energies and the photoexcited hole and electron wavefunctions were calculated using a constrained LSDA approach where the occupations of the conduction band and of the impurity level were fixed. For the case of the $ML_{CB}CT$ excited state, one electron was removed from the cobalt $e(\downarrow)$ orbital and transferred to the spin-down conduction band minimum. The transition energy is given by the total energy difference between the excited- and ground-state configurations. This methodology gives a constant shift in the Kohn-Sham eigenvalue energy differences over all nanocrystal sizes. This shift is related to electron-electron repulsion at the cobalt. Test calculations showed that the use of defect formation energies[35] gives similar results for the position of the transition energies. The partial densities of states were obtained by projecting the radial wavefunctions onto spherical harmonics inside a sphere of radius 2.4 a.u. centered at the atomic positions.

**References**

1.  Furdyna, J. K. & Kossut, J. in *Semiconductors and Semimetals* (eds. Willardson, R. K. & Beer, A. C.) (Academic Press, New York, 1988).
2.  Zutic, I., Fabian, J. & Das Sarma, S. Spintronics: Fundamentals and Applications. *Rev. Modern Phys.* **76**, 323-410 (2004).
3.  Jungwirth, T., Sinova, J., Masek, J., Kucera, J. & MacDonald, A. H. Theory of ferromagnetic (III,Mn)V semiconductors. *Rev. Modern Phys.*, in press (2006).
4.  Dietl, T., Ohno, H., Matsukura, F., Cibert, J. & Ferrand, D. Zener model description of ferromagnetism in zinc-blende magnetic semiconductors. *Science* **287**, 1019-1022 (2000).
5.  Wojtowicz, T., Kutrowski, M., Karczewski, G., Kossut, J., Konig, B., Keller, A., Yakovlev, D. R., Waag, A., Geurts, J., Ossau, W., Landwehr, G., Merkulov, I. A., Teran, F. J. & Potemski, M. II-VI Quantum Structures with Tunable Electron g-Factor. *J. Cryst. Growth* **214/215**, 378-386 (2000).
6.  Merkulov, I. A., Yakovlev, D. R., Keller, A., Ossau, W., Geurts, J., Waag, A., Landwehr, G., Karczewski, G., Wojtowicz, T. & Kossut, J. Kinetic Exchange between the Conduction Band Electrons and Magnetic Ions in Quantum-Confined Structures. *Phys. Rev. Lett.* **83**, 1431-1434 (1999).
7.  Mackh, G., Ossau, W., Waag, A. & Landwehr, G. Effect of the reduction of dimensionality on the exchange parameters in semimagnetic semiconductors. *Phys. Rev. B* **54**, R5227-R5230 (1996).
8.  Bhattacharjee, A. K. Confinement-induced reduction of the effective exchange parameters in semimagnetic semiconductor nanostructures. *Phys. Rev. B* **58**, 15660-15665 (1998).
9.  Sapra, S., Sarma, D. D., Sanvito, S. & Hill, N. A. Influence of quantum confinement on the electronic and magnetic properties of (Ga,Mn)As diluted magnetic semiconductor. *Nano Letters* **2**, 605-608 (2002).
10. Huang, X. Y., Makmal, A., Chelikowsky, J. R. & Kronik, L. Size-dependent spintronic properties of dilute magnetic semiconductor nanocrystals. *Phys. Rev. Lett.* **94**, 236801 (2005).
11. Adachi, S. & Taguchi, T. Optical properties of ZnSe. *Phys. Rev. B* **43**, 9596-9577 (1991).
12. Malikova, L., Krystek, W., Pollak, F. H., Dai, N., Cavus, A. & Tamargo, M. C. Temperature dependence of the direct gaps of ZnSe and $Zn_{0.56}Cd_{0.44}Se$. *Phys. Rev. B* **54**, 1819-1824 (1996).
13. Suyver, J. F., van der Beek, T., Wuister, S. F., Kelly, J. J. & Meijerink, A. Luminescence of Nanocrystalline ZnSe:Cu. *App. Phys. Lett.* **79**, 4222-4224 (2001).
14. Tsai, T. Y. & Birnbaum, M. $Co^{2+}$:ZnS and $Co^{2+}$:ZnSe saturable absorber Q switches. *J. Appl. Phys.* **87**, 25-29 (2000).
15. Ham, F. S., Ludwig, G. W., Watkins, G. D. & Woodbury, H. H. Spin Hamiltonian of $Co^{2+}$. *Phys. Rev. Lett.* **5**, 468-470 (1960).
16. Norberg, N. S., Parks, G. L., Salley, G. M. & Gamelin, D. R. Giant Excitonic Zeeman Splittings in $Co^{2+}$-doped ZnSe Quantum Dots. *J. Am. Chem. Soc.* **128**, in press (2006).
17. Norris, D. J., Yao, N., Charnock, F. T. & Kennedy, T. A. High-quality manganese-doped ZnSe nanocrystals. *Nano Letters* **1**, 3-7 (2001).
18. Madelung, O. *Semiconductors: Data Handbook* (Springer, Berlin, 2004).
19. Robbins, D. J., Dean, P. J., Glasper, J. L. & Bishop, S. G. New high-energy luminescence bands from $Co^{2+}$ in ZnSe. *Solid State Comm.* **36**, 61-67 (1980).
20. Noras, J. M., Szawelska, H. R. & Allen, J. W. Energy levels of cobalt in ZnSe and ZnS. *J. Phys. C* **14**, 3255-3268 (1981).

    

21. Sokolov, V. I., Surkova, T. P., Kulakov, M. P. & Fadeev, A. V. New Experimental Evidence Concerning the Nature of L, M, and N Lines in ZnSe:Co. *Phys. Stat. Solidi (b)* **130**, 267-272 (1985).

22. Ehlert, A., Dreyhsig, J., Gumlich, H.-E. & Allen, J. W. Excited-state absorption of ZnSe doped with cobalt. *J. Lumin.* **60-61**, 21-25 (1994).

23. Brus, L. Electronic Wave Functions in Semiconductor Clusters: Experiment and Theory. *J. Phys. Chem.* **90**, 2555-2560 (1986).

24. Kittilstved, K. R., Liu, W. K. & Gamelin, D. R. Electronic structure origins of polarity dependent high-$T_C$ ferromagnetism in oxide diluted magnetic semiconductors. *Nature Materials* **5**, 291-297 (2006).

25. Shklovskii, B. I. Hopping Conduction in Lightly Doped Semiconductors. *Sov. Phys. Semicond.* **6**, 1053-1075 (1973).

26. Berciu, M. & Bhatt, R. N. Effects of Disorder on Ferromagnetism in Diluted Magnetic Semiconductors. *Phys. Rev. Lett.* **87**, 107203 (2001).

27. Mott, N. Metal-Insulator Transitions. *Proc. Royal Soc. London, Series A* **382**, 1-24 (1982).

28. Langer, J. M., Delerue, C., Lannoo, M. & Heinrich, H. Transition-metal impurities in semiconductors and heterojunction band lineups. *Phys. Rev. B* **38**, 7723-7739 (1988).

29. Caldas, M. J., Fazzio, A. & Zunger, A. A universal trend in the binding energies of deep impurities in semiconductors. *App. Phys. Lett.* **45**, 671-673 (1984).

30. Van de Walle, C. G. & Neugebauer, J. Universal alignment of hydrogen levels in semiconductors, insulators and solutions. *Nature* **423**, 626-628 (2003).

31. Smith, C. A., Lee, H. W. H., Leppert, V. J. & Risbud, S. H. Ultraviolet-blue emission and electron-hole states in ZnSe quantum dots. *App. Phys. Lett.* **75**, 1688-1690 (1999).

32. Swank, R. K. Surface properties of II-VI compounds. *Phys. Rev.* **153**, 844-849 (1967).

33. Troullier, N. & Martins, J. L. Efficient pseudopotentials for plane-wave calculations. *Phys. Rev. B* **43**, 1993-2006. (1991).

34. Chelikowsky, J. R., Troullier, N. & Saad, Y. Finite-difference-pseudopotential method: electronic structure calculations without a basis. *Phys. Rev. Lett.* **72**, 1240-1243 (1994).

35. Dalpian, G. M. & Wei, S.-H. Electron-induced stabilization of ferromagnetism in $Ga_{1-x}Gd_xN$. *Phys. Rev. B* **72**, 115201 (2005).

**Acknowledgments.** Financial support from the US NSF (DMR-0239325), Research Corporation, the Dreyfus Foundation, and the Sloan Foundation is gratefully acknowledged. Work at UT Austin was supported by the US NSF (DMR-0551195) and the US DOE (DE-FG02-89ER45391 and DE-FG02-03ER15491). Calculations were performed at TACC and NERSC.

**Correspondence.** Gamelin@chem.washington.edu

**Fig. 1. Spectroscopic results for $Co^{2+}$:ZnSe quantum dots. (A)** 300 K electronic absorption spectra of colloidal 5.6 nm diameter 0.77% $Co^{2+}$:ZnSe nanocrystals. The dotted line is the $Co^{2+}$ $^4A_2(F) \rightarrow {}^4T_1(P)$ absorption band of bulk ≤0.1% $Co^{2+}$:ZnSe.[14] **(B)** 5 K electronic absorption and **(C)** 5 K, 6 T MCD spectra of $Co^{2+}$:ZnSe QDs of 4.1, 4.6, and 5.6 nm diameters (0.61, 1.30, 0.77% $Co^{2+}$, respectively). **Inset:** Schematic illustration of (*i*) ligand-field (LF), (*ii*) charge-transfer (CT), and (*iii*) excitonic (EXC) transitions.

**Fig. 2. Analysis of charge transfer energies. (A)** The CT MCD spectra from Fig. 1C plotted on an expanded scale. **(B)** The experimental CT (●) and excitonic (♦) transition energies for the $Co^{2+}$:ZnSe QDs from Fig. 1 plotted vs the excitonic energy. The best fit line to the CT data (dashed green) has slope 0.80 ± 0.03, and its extrapolation yields a predicted CT energy of 2.51 ± 0.07 eV for bulk $Co^{2+}$:ZnSe (○). The red bars represent low-temperature $ML_{CB}CT$ energies reported for bulk $Co^{2+}$:ZnSe.[19-22] **(C)** Experimental and **(D)** calculated valence and conduction band energies and the pinned $Co^{2+}$ energies in $Co^{2+}$:ZnSe, plotted vs ZnSe nanocrystal diameter. The arrows represent the $ML_{CB}CT$ (solid) and $L_{VB}MCT$ (dashed) transition energies. The energies in (C) are plotted relative to vacuum based on the experimental valence-band ionization energy for bulk ZnSe. The difference between the VB and $Co^{2+}/Co^{3+}$ level is the $Co^{2+}$-hole binding energy, $E_b$, of the $ML_{CB}CT$ excited state (see text).

**Fig. 3. Results from DFT/LSDA calculations on $Co^{2+}$:ZnSe QDs.** Probability density of **(A)** the photoexcited electron and **(B)** the photogenerated hole in the $ML_{CB}CT$ excited state of the 1.45 nm diameter (87 atom) nanocrystal. **(C,D)** Density-of-states (DOS) plots for the $ML_{CB}CT$ excited states of the 1.11 and 2.27 nm diameter nanocrystals. The red curves are the partial $Co(3d)$ contributions to the total DOS. The energy difference between the $e(\downarrow)$ orbital and the valence band edge increases with increasing quantum confinement. Energies are plotted relative to $e(\downarrow)$.

**Fig. 4. Quantum confinement and impurity energy levels. (A)** The effective Bohr radius for the hole bound to $Co^{2+}$ in the $Co^{2+}$:ZnSe $ML_{CB}CT$ excited state (dashed) and to $Mn^{2+}$ in the $Ga_{1-x}Mn_xAs$ ground state (solid), calculated from Eq. 3 and plotted vs particle diameter. **(B)** The critical manganese acceptor concentration necessary for the metal-insulator transition in $Ga_{1-x}Mn_xAs$ plotted vs particle diameter, calculated using Eqs. 1, 3, and 4.

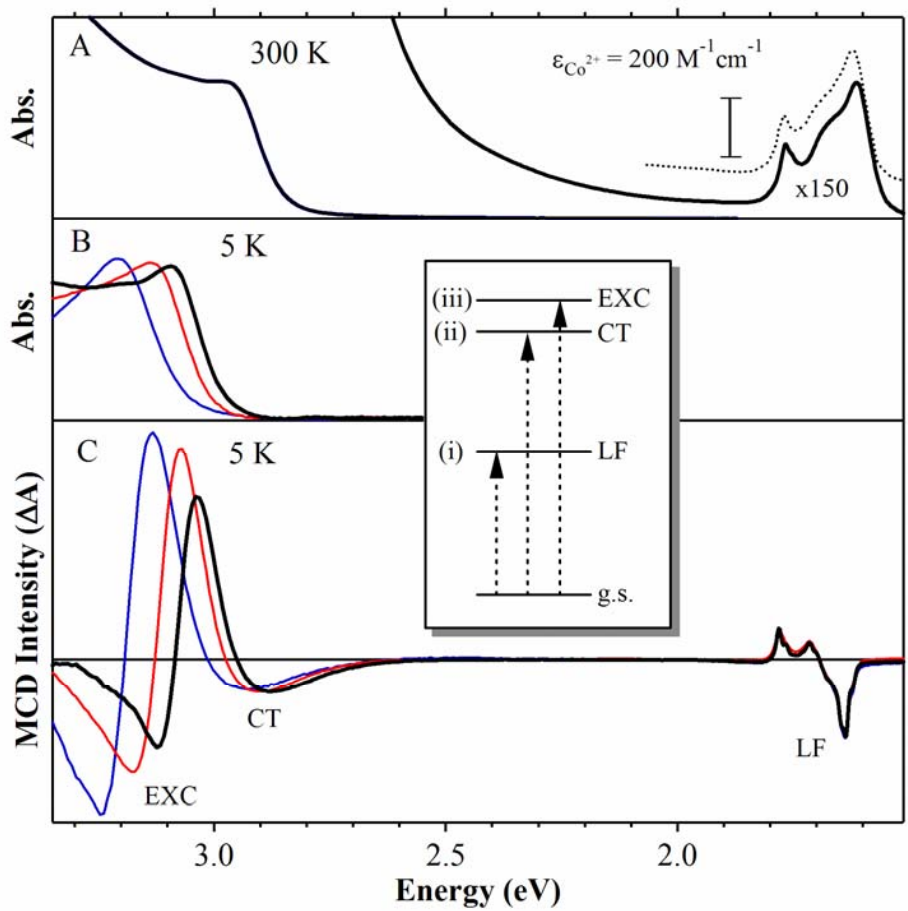

**Figure 1.**

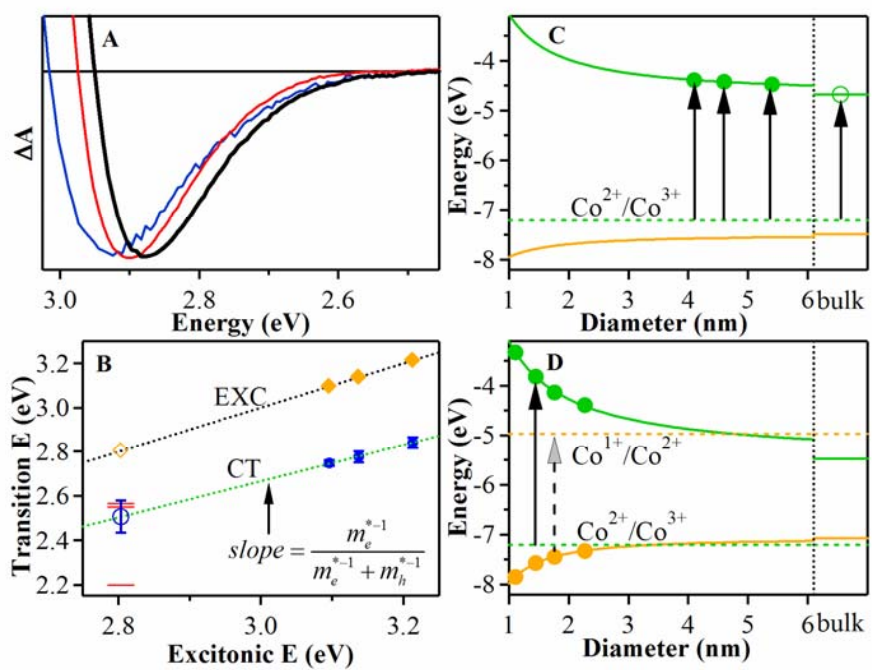

**Figure 2.**

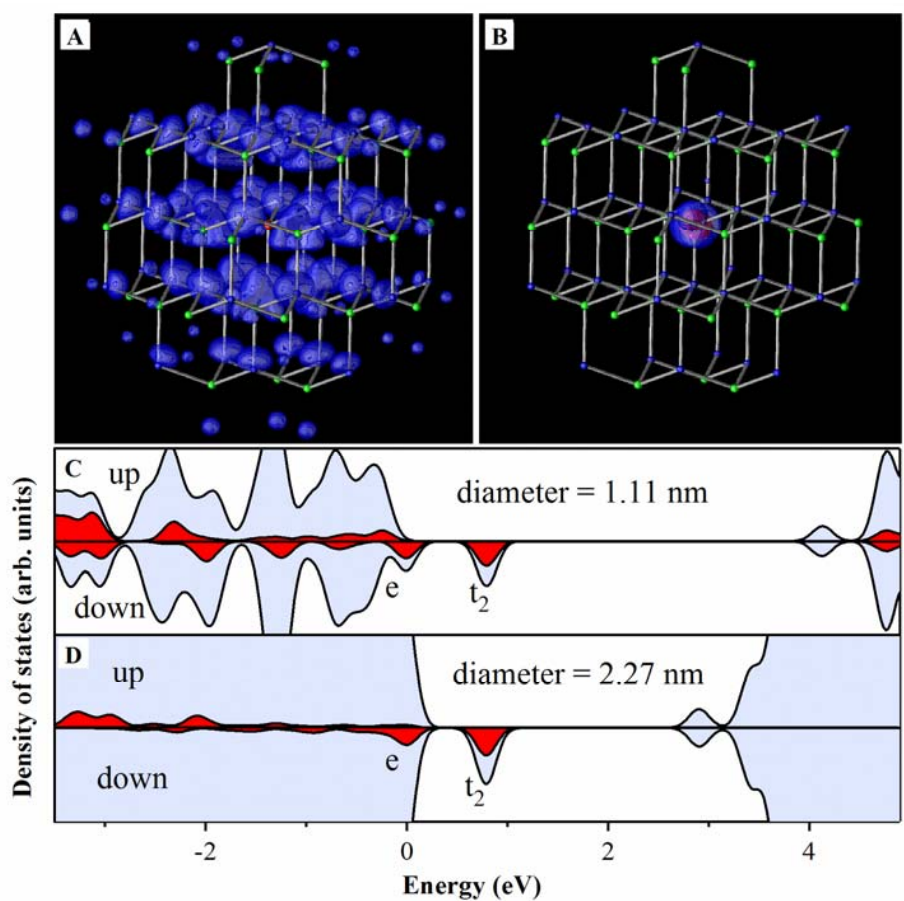

**Figure 3.**

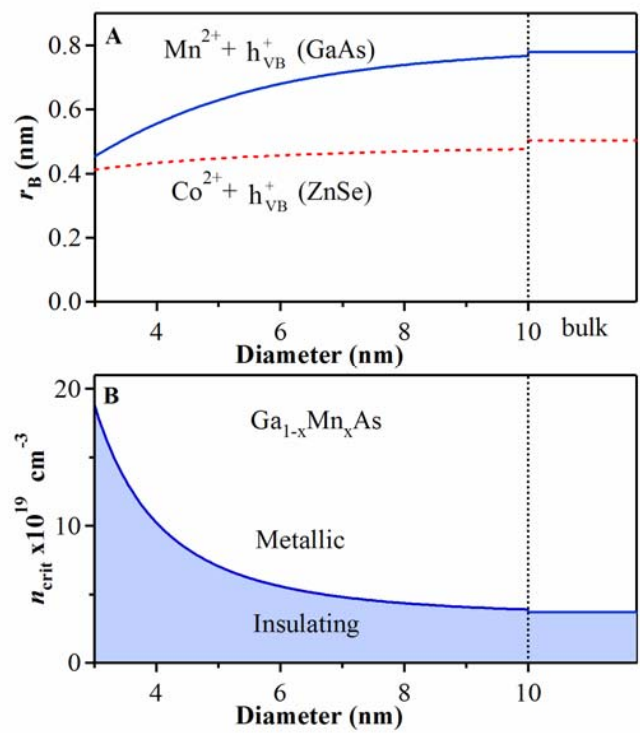

**Figure 4.**